\documentclass[twocolumn,showpacs,preprintnumbers,amsmath,amssymb,floatfix]{revtex4} 

\usepackage{graphicx}
\usepackage{dcolumn}
\usepackage{bm}

\begin{document}

\title{Sympathetic Cooling of Trapped Cd$^+$ Isotopes}

\author{B.B. Blinov, L. Deslauriers, P. Lee, M.J. Madsen, R. Miller and C. Monroe}
\address{University of Michigan Department of Physics\\ and FOCUS Center\\
		Ann Arbor, MI 48109-1120, USA}

\date{\today}

\begin{abstract}
We sympathetically cool a trapped $^{112}$Cd$^{+}$ ion by directly Doppler-cooling a $^{114}$Cd$^{+}$
ion in the same trap. This is the first demonstration of optically addressing a single trapped ion
being sympathetically cooled by a different species ion. 
Notably, the experiment uses a single laser source, and does not require strong focusing.
This paves the way toward reducing decoherence in an ion trap quantum computer
based on Cd$^+$ isotopes.
\end{abstract}

\pacs{03.67.Lx, 32.80.Pj, 42.50.Vk}

\maketitle

A collection of cold trapped ions offers one of the most promising avenues towards 
realizing a quantum computer~\cite{trappedions,monroe,sorensen,wine1}. Quantum information is stored
in the internal states of individual trapped ions, while entangling quantum logic
gates are implemented via a collective quantized mode of motion of the ion crystal.
The internal qubit states can have extremely long coherence times~\cite{bollinger}, 
but decoherence of the motion of the ion
crystal may limit quantum logic gate fidelity~\cite{decoherence}. Furthermore, when ions
are nonadiabatically shuttled between different trapping regions
for large-scale quantum computer schemes~\cite{wine1,ccd}, their motion
must be recooled for subsequent logic operations.

Direct laser cooling of the qubit ions is not generally possible without disturbing
coherence of the internal qubit states. Instead, additional ``refrigerator" ions in the crystal
can be directly laser-cooled, with the qubit ions cooled in sympathy by virtue of their
Coulomb-coupled motion~\cite{larson}. The laser cooling of the refrigerator ions can quench unwanted
motion of the ion crystal~\cite{wine1,symcool}, while not affecting the internal states of the qubit ions. 
Sympathetic 
cooling has been observed in large ensembles of ions in Penning traps~\cite{larson,imajo},
impurities in small collections of ion crystals, and in small ion crystals consisting
of a single species, where strong laser focusing was required to access a particular ion without 
affecting the others~\cite{rhode}. 
Here, we report the first demonstration of sympathetic
cooling in a small ion crystal with two different species where both species are 
independently optically addressed.

We study sympathetic cooling of Cd$^+$ isotopes in an asymmetric-quadrupole rf trap.
One ion isotope (the refrigerator ion) is
continuously Doppler-cooled by a laser beam red-detuned from its D2 line (S$_{1/2}$-P$_{3/2}$), 
while the other isotope (the probe ion) is either Doppler-cooled 
or Doppler-heated by another beam, whose frequency is scanned around its D2 resonance line.
The effect of the sympathetic cooling is to enable measuring fluorescence on the blue side of the probe
ion's resonance. 
Ordinarily, when the probe laser beam is tuned to the blue of the probe ion's resonance, 
the ion ceases  fluorescing due to Doppler-heating, but the sympathetic cooling from the
refrigerator ion keeps the probe ion cold and fluorescing regardless of the probe tuning.

In the experiment, the probe ion is $^{112}$Cd$^{+}$, while the refrigerator ion is $^{114}$Cd$^{+}$
(both isotopes have zero nuclear spin). 
The respective D2 lines of these two neighboring isotopes are separated by
about 680~MHz, with the heavier ion at lower frequency, 
and each ion's natural width is $\gamma/2\pi\simeq47$~MHz.

\begin{figure}
\includegraphics{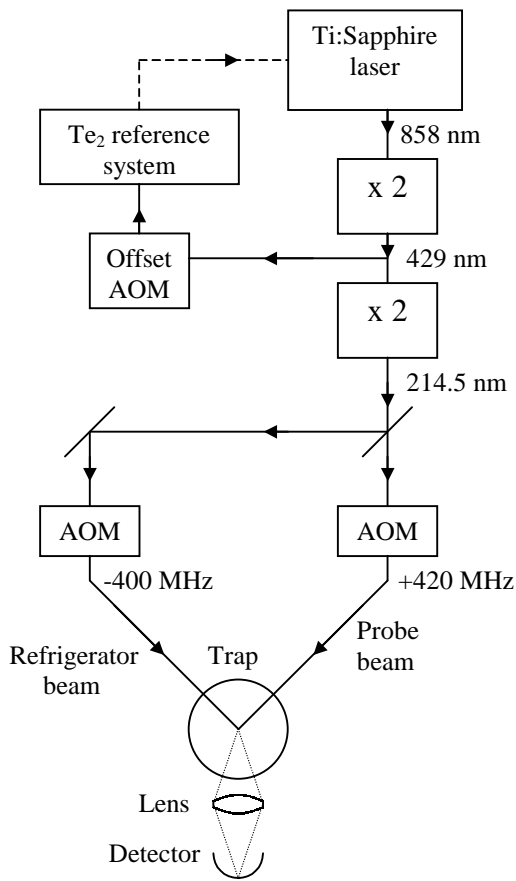}
\caption{Schematic diagram of the experiment. The 858~nm light from the Ti:Sapphire laser
is frequency doubled, then a small portion ($\sim$20~mW) of the 429~nm light is diverted to a Tellurium 
saturated absorption spectrometer. The double-pass acusto-optic modulator (AOM) in the Te$_2$ reference system allows 
tuning the Ti:Sapphire
beam frequency by about $\pm$25~MHz while locked. The remaining ($\sim$200~mW) portion of the
429~nm light is again frequency doubled to $\sim$15~mW of 214.5~nm radiation.
The UV beam is then split into two parts, each of which is frequency-shifted by
AOMs and directed into the trap.}
\end{figure}

The experimental apparatus is schematically shown in Fig.~1. The Cd$^+$ D2 line resonant 
light near 214.5~nm is generated by quadrupling a Ti:Sapphire laser. 
The laser is stabilized to a molecular Tellurium feature near 429~nm to better than 1~MHz.
The quadrupled UV output is split into two parts; one part is
upshifted by $\sim$420~MHz, while the other downshifted by $\sim$400~MHz using 
acusto-optical frequency shifters.
The two beams are then directed into the ion trap through separate windows. 
Both beams uniformly illuminate the ion crystal in the trap.
The up-shifted UV beam (the probe beam) 
is scanned in frequency around the $^{112}$Cd$^{+}$ ion's D2 line, while the down-shifted UV beam
(the refrigerator beam) frequency is
always kept to the red of the $^{114}$Cd$^{+}$ ion's D2 line. 
The UV fluorescence from the ions is collected by an f/5.6 lens and 
imaged onto a micro-channel plate detector. 
The fluorescence counts are integrated for 10~s for each data point in a frequency scan.

We use an asymmetric quarupole RF ion trap~\cite{jeffers} with the electrode radius of about 200 microns. 
The trap's RF frequency is $\Omega/2\pi\simeq38.8$~MHz, the RF potential amplitude is about 200~V,
and the trap's electrodes are kept at the same static potential.
The measured secular trap frequency along the ions' separation axis is $\omega_x/2\pi\sim2.8$~MHz. 

To study sympathetic cooling, we load a $^{112}$Cd$^{+}$ ion and a $^{114}$Cd$^{+}$ ion
into the trap by ablating and photoionizing the neutral Cd deposited on the trap's electrodes
using one of the UV beams.
Due to high abundance of isotopes $^{112}$Cd and $^{114}$Cd in natural cadmium 
(24\% and 29\%, respectively), loading the proper two isotopes is not unlikely. We set the trap's
compensating electrode voltages such that the $^{112}$Cd$^{+}$ (probe) ion is near the rf null
of the trap to minimize its micromotion and thus simplify its lineshape~\cite{micromotion}.

In Fig.~2, the fluorescence from the $^{112}$Cd$^{+}$ ion is plotted against the 
probe beam frequency.
In Fig.~2a, both the probe and the refrigerator
laser beams are on, while for the data in Fig.~2b 
the refrigerator beam is turned off. 
Note the telltale drop in fluorescence as the 
probe beam is tuned to the blue side of the $^{112}$Cd$^{+}$ resonance line in Fig.~2b caused by 
Doppler-heating. On the other hand, the fluorescence 
curve in Fig.~2a is symmetric, which demonstrates that the $^{112}$Cd$^{+}$ ion is
sympathetically cooled by the $^{114}$Cd$^{+}$ ion even as the probe beam is tuned to the blue
of the resonance.

\begin{figure}[b]
\includegraphics{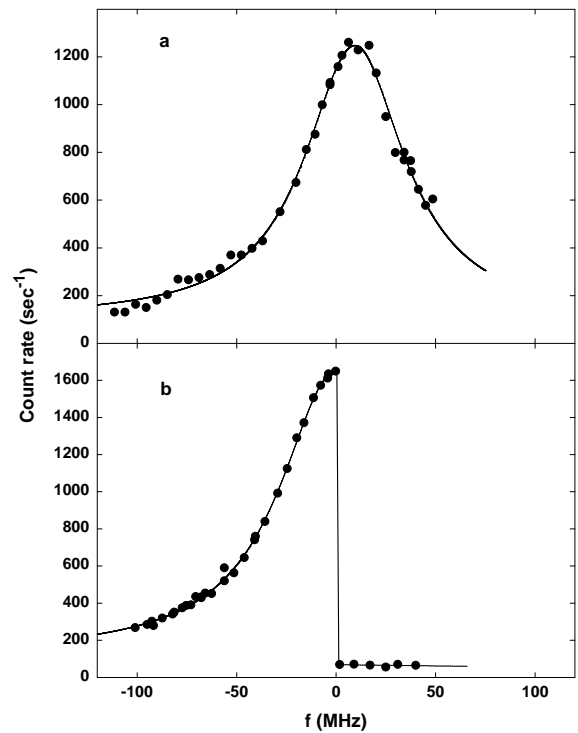}
\caption{UV fluorescence count rate from the $^{112}$Cd$^{+}$ probe ion (a) with and (b) without sympathetic cooling,
plotted against the probe beam frequency detuning from resonance. 
Solid lines represent fits to the data using: (a) a Voigt profile,
and (b) a Voigt shape for the below-resonance part of the data and a straight line for above-resonance
part of the data.}
\end{figure}

Images of the two ions at different lighting conditions are shown in Fig.~3; the probe ion ($^{112}$Cd$^+$) is on the
left, while the refrigerator ion ($^{114}$Cd$^+$) is to the right. Both the probe and the refrigerator
beams are turned on for Fig.~3a; in Fig.~3b, only the probe beam is on, while in Fig.~3c only the refrigerator
beam is on. Note a very faint images of the $^{114}$Cd$^{+}$ ion in Fig.~3b 
and the $^{112}$Cd$^{+}$ ion in Fig.~3c from the
residual fluorescence from the far-detuned beams.

\begin{figure}
\includegraphics{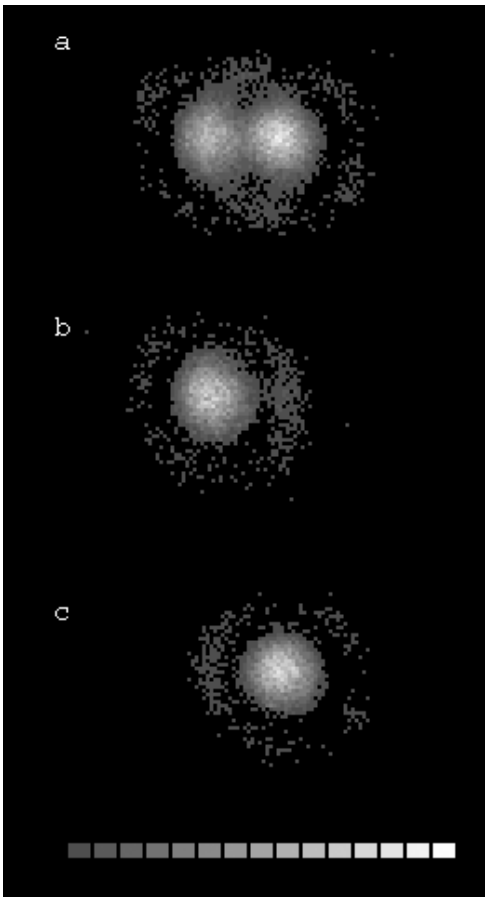}
\caption{Images of the $^{112}$Cd$^{+}$ and $^{114}$Cd$^{+}$ ions while illuminated by (a)
both the refrigerator and the probe beam, (b) only the probe beam and (c) only the refrigerator
beam. The $^{112}$Cd$^{+}$ ion is on the left, while the $^{114}$Cd$^{+}$ ion is on the right.
The two ions are separated by $\sim$2~$\mu$m.
Note very faint images of the $^{114}$Cd$^{+}$ ion in (b) 
and the $^{112}$Cd$^{+}$ ion in (c) from the
residual fluorescence from the far-detuned beams.
The below scale is linear in the integrated photon counts. The exposure time is
10~s for each picture.}
\end{figure}

For the data shown in Figs. 2 and 3, the probe beam intensity is $I_{probe} = 0.35I_{sat}$,
while the refrigerator beam intensity is $I_{ref} = 12I_{sat}$, where the saturation intensity
$I_{sat}\simeq 0.6$~W/cm$^2$. Such high refrigerator beam intensity is necessary because of the large amount 
of refrigerator ion micromotion. This significantly Doppler-broaden the refrigerator ion lineshape~\cite{micromotion} and
makes cooling less efficient than the comparable cooling/heating of the probe ion, which experiences little
micromotion. In a linear ion trap, where micromotion can be suppressed in all ions in a string, 
the required cooling intensity would not need to be as high.

To demonstrate that the cooling seen in Fig.~2a is not caused by directly Doppler-cooling of 
the $^{112}$Cd$^{+}$ probe ion
by the refrigerator beam (which indeed is red-detuned from the $^{112}$Cd$^{+}$ ion's D2 line) we load
a single $^{112}$Cd$^{+}$ ion into the trap, while shining both the refrigerator and the probe beams onto the ion.
The curve in Fig.~4 shows the resulting fluorescence as function of the probe laser frequency. When tuned to the
blue of the resonance, the ion's fluorescence quickly drops to zero, indicating that the ion is heated by the
probe beam; the direct Doppler-cooling by the far-detuned refrigerator beam is not sufficient to keep the ion cold.

\begin{figure}
\includegraphics{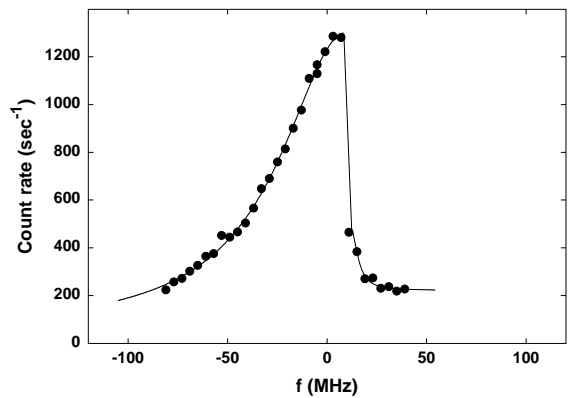}
\caption{UV fluorescence from a single $^{112}$Cd$^{+}$ ion while illuminated by both the refrigerator
and the probe beam plotted against the probe beam frequency detuning from the resonance. 
The solid line is a Voigt fit to the data
below the resonance and a hand-drawn line above the resonance.}
\end{figure}

While the current experiment only investigates Doppler cooling, quantum logic gates with trapped 
ions generally require cooling to the Lamb-Dicke limit, where the ion's 
spatial extent is much smaller than the optical coupling wavelength. 
A similar setup should enable sympathetic cooling of Cd$^+$ isotopes
to the Lamb-Dicke limit using stimulated-Raman sideband cooling~\cite{raman}, polarization-gradient
(Sisyphus) cooling~\cite{sisyphus}, or EIT cooling~\cite{eit}.

Ultimately, we plan to sympathetically cool $^{111}$Cd$^{+}$ qubits 
(nuclear spin-1/2) with $^{116}$Cd$^{+}$ refrigerator ions.
The isotope shift between Cd$^+$ isotopes 111 and 116 is $\Delta/2\pi\simeq5.2$~GHz.
Below, we estimate the decoherence of a $^{111}$Cd$^{+}$ qubit under the influence of 
$^{116}$Cd$^{+}$ cooling radiation, assuming a background heating source is quenched
by sideband cooling in the Lamb-Dicke limit. The photon scatter rate
by the $^{116}$Cd$^{+}$ refrigerator ion $\Gamma_s\simeq\frac{I}{I_{sat}}\gamma/2$
sets an upper limit on the cooling rate, which itself must be at least as large
as the heating rate $\dot{n}$ for useful sympathetic cooling.
The off-resonant spontaneous emission rate of the $^{111}$Cd$^{+}$ qubit ion 
is then of order $\Gamma_{qubit}\simeq\dot{n}\frac{\gamma^2}{4\Delta^2}\simeq0.02$/sec
for a heating rate of $\dot{n}\simeq10^3$/sec~\cite{decoherence}.
Under the same conditions, the AC Stark shift of the $^{111}$Cd$^{+}$ qubit ion is
$\delta_{AC}/2\pi\simeq\frac{\dot{n}}{2\pi}\frac{\gamma}{4\Delta}\simeq0.3$~Hz, and only the fluctuations
of this already small shift will cause decoherence. A similar analysis can be given 
for other methods of cooling.
We find that the 5.2~GHz isotope shift appears large enough to comfortably neglect
qubit decoherence from spontaneous emission and AC Stark shifts, while it is
small enough so that optical modulators can provide the cooling radiation
without the need for additional laser sources.

In summary, we have sympathetically cooled a single trapped $^{112}$Cd$^{+}$ ion through
Doppler-cooling of a neighboring $^{114}$Cd$^{+}$ ion. This is the first demonstration
of optically addressing a single trapped ion being sympathetically cooled by an ion of
a different species. The sympathetic cooling of multiple ion species is an important
step toward scaling the trapped ion quantum computer, as it can reduce decoherence
associated with unwanted motion of trapped ions, while preserving the internal qubit
coherence. The Cd$^+$ system is convenient, as the sympathetic cooling can be accomplished without
extra lasers and without strong focusing.

We wish to thank J. Zorn, R. Conti for their useful advice, and R. Jennerich for his help with the earlier
parts of the experiment. This work is supported by
the U.S. National Security Agency, the Advanced Research and Development Activity,
the Army Research Office, and the National Science Foundation.

\end{document}